\documentclass{article}
\usepackage[preprint]{spconfa4}
\usepackage{enumitem}
\usepackage{amsmath,epsfig}
\usepackage{graphicx,url}
\usepackage{soul}
\usepackage{inputenc}  
\usepackage{verbatim}
\usepackage{listings}
\usepackage{indentfirst}
\usepackage{booktabs,multirow}
\usepackage{tikz}
\usepackage{textcomp}
\usepackage{flushend}

\let\OLDthebibliography\thebibliography
\renewcommand\thebibliography[1]{
  \OLDthebibliography{#1}
  \setlength{\parskip}{0pt}
  \setlength{\itemsep}{0pt plus 0.3ex}
}

\begin{document}\sloppy
\ninept
\def\x{{\mathbf x}}
\def\L{{\cal L}}

\title{Encoding in the Dark Grand Challenge: An Overview}
%
\name{Nantheera Anantrasirichai, Fan Zhang, Alexandra Malyugina, Paul Hill, and Angeliki Katsenou}
\address{Visual Information Laboratory, University of Bristol, UK}

\maketitle

\begin{abstract}
A big part of the video content we consume from video providers consists of genres featuring low-light aesthetics. Low light sequences have special characteristics, such as spatio-temporal varying acquisition noise and light flickering, that make the encoding process challenging. To deal with the spatio-temporal incoherent noise, higher bitrates are used to achieve high objective quality. Additionally, the quality assessment metrics and methods have not been designed, trained or tested for this type of content. This has inspired us to trigger research in that area and propose a Grand Challenge on encoding low-light video sequences. In this paper, we present an overview of the proposed challenge, and test state-of-the-art methods that will be part of the benchmark methods at the stage of the participants' deliverable assessment. From this exploration, our results show that VVC already achieves a high performance compared to simply denoising the video source prior to encoding. Moreover, the quality of the video streams can be further improved by employing a post-processing image enhancement method. 
\end{abstract}
\vspace{-2mm}
\section{Introduction}
\label{sec:intro}
\vspace{-1mm}
\noindent Last year, the HBO Game of Thrones episode, entitled ``The Long Night'', received a lot of controversial reviews because it was shot in low light and many fans complained about the picture quality \cite{w:GameofThrones}. Low light scenes often come with acquisition noise, which not only disturbs the viewers, but also brings special characteristics to video compression. This noise appears randomly in time, possibly creating visibly temporal flickering. These type of videos are often encountered in cinema as a result of artistic perspective or the nature of a scene. Other examples include shots of wildlife (e.g.\ mobula rays at night in Blue Planet II), concerts, shows, surveillance camera footage and more. In this context, we  study the encoding of videos captured in low-light using state-of-the-art methods that has inspired us to organise a video coding Grand Challenge within IEEE ICME2020. 

Noise can be introduced during video acquisition, recording and processing. Not only visually unpleasant, noise also affects the performance of intra and inter prediction in video compression, causing the encoder to inefficiently spend bits to represent this noise, especially at low compression levels. Currently, noise reduction techniques are usually employed for film-grain noise in the creative industry during a pre-encoding phase with the aim of improving compression performance. Later, synthetic noise is superimposed at the decoded video sequence~\cite{Stankiewicz:video:2019,KuoTCSVT2009}. However, film-grain noise is a special case as it is considered part of the artistic effect that enhances the natural appearance of the video and the viewers are quite comfortable with it. On the other hand, there are other types of noise that the viewers do not like and perceive in many cases as `low' quality. In this paper, we consider types of noise that are undesirable for viewers, consequently, we do not attempt to restore the noise after the decoding. 
The simplest technique of noise reduction is a weighted averaging technique performed in a temporal sliding window---a.k.a. moving average filter~\cite{Yahya:video:2016}. More sophisticated methods include adaptive spatio-temporal smoothing through anisotropic filtering~\cite{Malm:adaptive:2007}, nonlocal transform-domain group filtering~\cite{Maggioni:BM4D:2012}, Kalman-bilateral mixture model~\cite{Zuo:video:2013}, and spatio-temporal patch-based filtering~\cite{Buades:CFA:2019}.
Recent work employs the popular deep learning approach.  For example, a residual noise map is estimated in the Denoising Convolutional Neural Network (DnCNN) method~\cite{Zhang:DnCNN:2017} for image based denoising, and for video based denoising, a spatial and temporal network are concatenated where the latter handles brightness changes and temporal inconsistencies~\cite{Claus:ViDeNN:2019}. VNLnet combines a non-local patch search module with DnCNN. The first part extracts features, and the latter mitigates the remaining noise~\cite{Davy:nonlocal:2019}.

Another direction in video coding is to perform denoising in the loop, such as~\cite{KaupCSVT2014,YangICASSP2017}. In-loop filters have been proposed and adopted by recent video coding standards (e.g.~\cite{JVET-O0079}). The most popular examples of in-loop filters are the adaptive deblocking, the adaptive loop, the sample adaptive offset, and Convolutional Neural Network (CNN) based filters. These filters however are meaningful in light compression, where they contribute towards better intra prediction. In heavy compression, quantization filters out the noise~\cite{KaupCSVT2014}. CNN-based methods are also popular for postprocessing and can provide significant image enhancement, leading to better final rate-distortion performance with significantly lower complexity~\cite{MaAccess2019, Zhang:Enhancing:2020}. 

In this paper, we briefly describe the Grand Challenge 
we are organising on encoding low-light sequences and we present some solutions using state-of-the-art methods to improve the compression performance. As quality metrics play an important role in the assessment of both
anchors and the challenge participant submissions, we explore the 
performance of state-of-the-art objective Image and Video Quality Assessment (IQA/VQA) metrics on low-light video content.
We first test the performance of recent video codec standard---Versatile Video Coding (VVC) on this type of content. Subsequently, we implement and present two possible workflows combining the codec and the denoising module: i) applying denoising before coding (pre-processing methods), and ii) applying image enhancement after coding (post-processing methods). 

The remainder of this paper is organised as follows. Section~\ref{sec:dataset} discusses the difficulties of encoding dark-scene videos. Then the evaluation metrics and the benchmark methods are described in Section~\ref{sec:metrics} and Section~\ref{sec:methods}, respectively. Section~\ref{sec:results} presents the evaluation results with discussion. Finally, a conclusion is outlined in Section~\ref{sec:conclusion}.

\vspace{-2mm}
\section{Low light videos and the difficulty in encoding}
\label{sec:dataset}
\vspace{-1mm}
\noindent Inspired by all above, we are organising a challenge on encoding low-light captured videos within IEEE ICME 2020. This challenge intends to identify technology that improves the perceptual quality of compressed low-light videos beyond the current state-of-the-art performance of the most recent coding standards, such as HEVC, AV1, VVC, etc. The challenge encourages the exploration of novel technologies applied within existing video codecs (e.g.\ in loop filters), or/and prior to encoding processing methods (e.g.\ denoising) and/or post-processing methods (e.g.\ de-blocking).

\begin{figure}[t!]
	\centering
  		\includegraphics[width=\columnwidth]{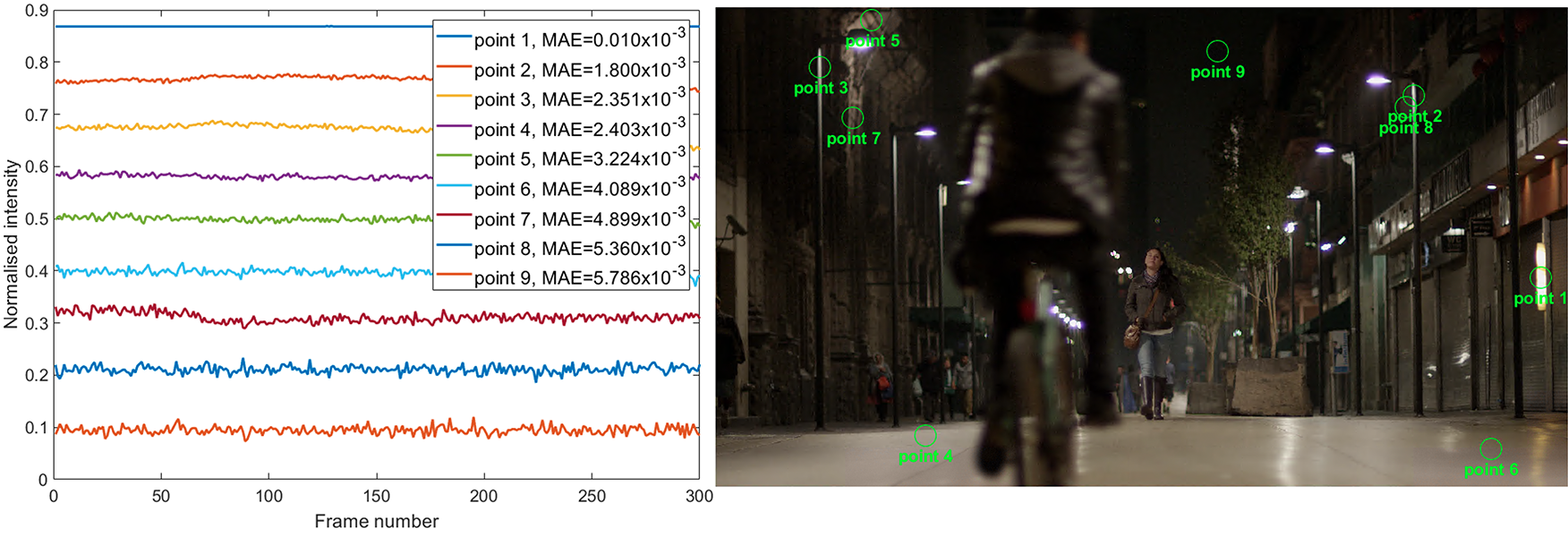}
    \vspace{-.5cm}
	\caption{Normalised intensity values (left) at various background locations (right) in each time frame of sequence S3.}
	\vspace{-.3cm}
	\label{fig:temporalsignals}
\end{figure}

The darker the light conditions, the harder it is to effectively capture video.  However, such conditions can be very aesthetically and artistically rewarding.  Moreover, shooting with low light also means that ISO noise will be present and it will be temporally incoherent. Frequently, light sources are not always consistent throughout the shot, posing additional challenges.
To illustrate an example of the temporal variation of intensity values in a dark sequence, we plotted the Y component as a time series for different intensity values in Fig.~\ref{fig:temporalsignals}. 
We computed the mean absolute errors (MAEs) from the smooth curves (an average of a sliding window with the size of 20 frames), and found that the darker areas have larger MAEs, implying higher noise levels. The brighter pixels have lower variance, but there is still a small temporal variation  despite the fact that all points are part of a static background. This is because the dark scenes generally contain several different light sources. Although this temporal variation may be unnoticeable to the viewers, it could significantly deteriorate the encoding performance and make a high target quality very expensive in terms of bitrate.

\vspace{-2mm}
\subsection{Dataset}
\begin{figure}[t!]
\scriptsize
\centering
\begin{minipage}[b]{0.325\linewidth}
\centering
\centerline{\includegraphics[width=1\linewidth]{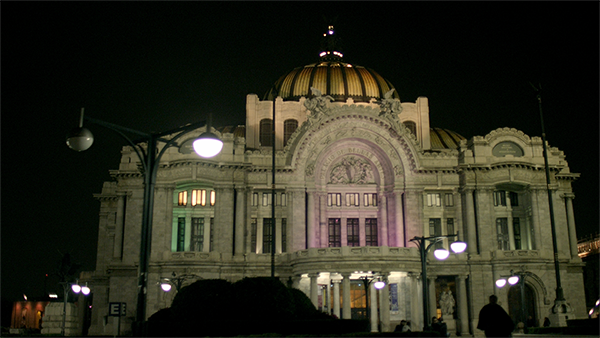}}
S1: ElFuente-Palacio\cite{w:NetflixElFuente}
\end{minipage}
\begin{minipage}[b]{0.325\linewidth}
\centering
\centerline{\includegraphics[width=1\linewidth]{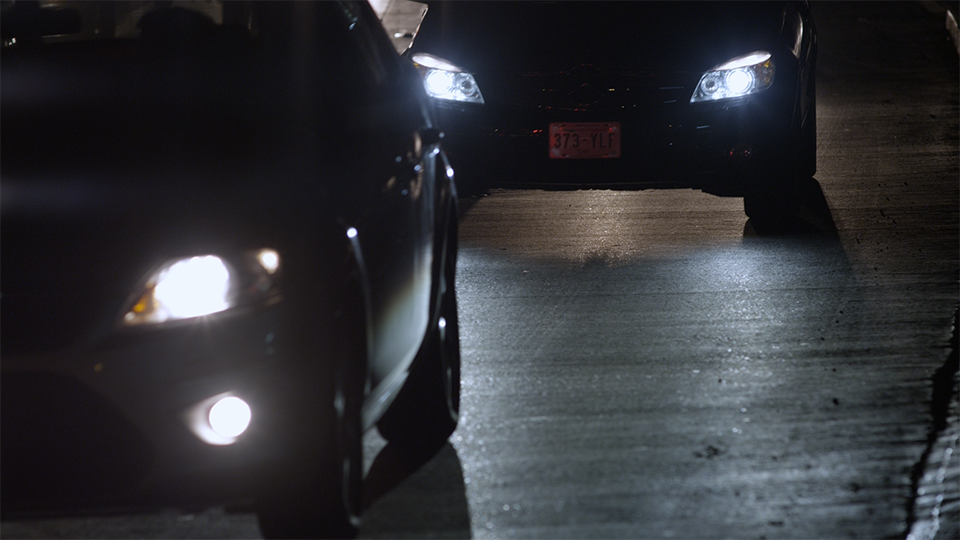}}
S2: ElFuente-Cars\cite{w:NetflixElFuente}
\end{minipage}
\begin{minipage}[b]{0.325\linewidth}
\centering
\centerline{\includegraphics[width=1\linewidth]{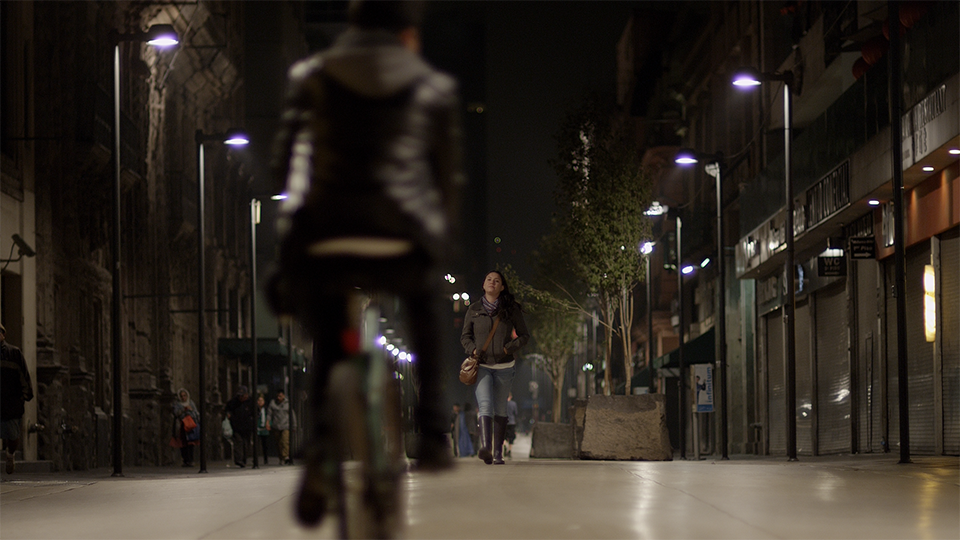}}
S3: ElFuente-Cyclist\cite{w:NetflixElFuente}
\end{minipage}
\begin{minipage}[b]{0.325\linewidth}
\centering
\centerline{\includegraphics[width=1\linewidth]{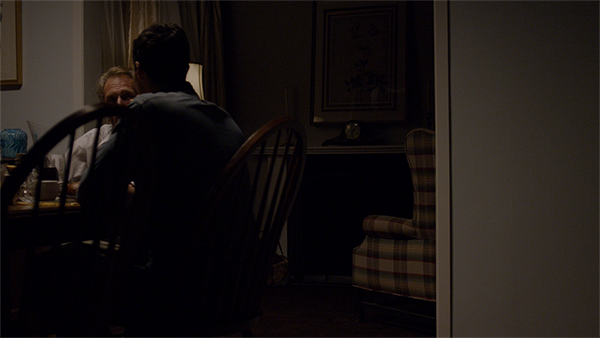}}
S4: Chimera-Dinner\cite{w:NetflixChimera}
\end{minipage}
\begin{minipage}[b]{0.325\linewidth}
\centering
\centerline{\includegraphics[width=1\linewidth]{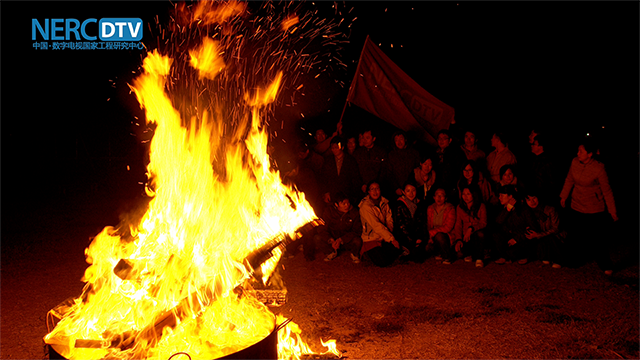}}
S5: Campfire\cite{c:SJTU4K}
\end{minipage}
\begin{minipage}[b]{0.325\linewidth}
\centering
\centerline{\includegraphics[width=1\linewidth]{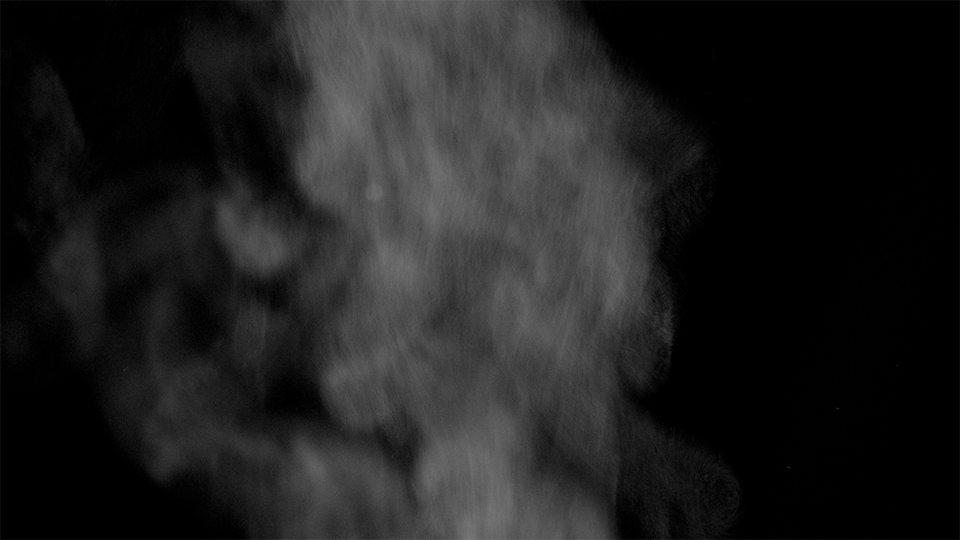}}
S6: SmokeClear\cite{Papadopoulos2015}
\end{minipage}
\vspace{-.1cm}
\caption{Sample frames from the test dataset.}
\vspace{-.5cm}
\label{fig:TestSeq}
\end{figure}

\vspace{-1mm}
\noindent For the Grand Challenge dataset\footnote{https://ieee-dataport.org/competitions/encoding-dark}
, we have selected 6 Full High Definition (FHD) (1920$\times$1080) YUV sequences with 10 bits depth and 420 colour sampling. All of the sequences except for Campfire (30fps) are at 60fps. The detailed list and sources of the test dataset is reported in Fig~\ref{fig:TestSeq}. We have carefully selected the sequences of different content, namely to include static or moving scenes, different types of motion, areas of interest, different luminance distribution including one or more light sources, varying background, and also dynamic textures (S5-S6).  
Sequences S1-S5 were captured by professionals, all are free for research purposes and all have been used in video coding standardisation activities or recent publications. Sequence S6 was captured by University of Bristol for a study of dynamic textures. 

\vspace{-2mm}
\subsection{Video Coding}
\label{ssec:VVC}
\vspace{-1mm}
\noindent Since its launch in 2004, H.264/AVC~\cite{r:h264} has been the most deployed video coding standard, despite the fact that its successor, H.265/HEVC~\cite{r:HEVC,b:Wien,j:Ohm} released in 2013, as it provides enhanced coding performance. The MPEG standardisation body is currently working towards the next generation video coding standard, Versatile Video Coding (VVC). It has been named versatile, as it is supporting immersive formats (higher spatial resolutions, higher dynamic range and 360\textdegree~videos). 
Recently, there has been increased activity in the video coding technology industry with the aim to develop open-source royalty-free video codecs, particularly by the Alliance for Open Media (AOMedia). In 2018, AV1 (AOMedia Video 1)~\cite{w:AV1} was released as a competitor to HEVC. AV1 was primarily based on Google's video codec VP9~\cite{w:VP9} and has comparable performance to HEVC~\cite{c:Zhang24, c:Topiwala, c:Grois,  c:Nguyen1}. 


\vspace{-2mm}
\section{Evaluation Metrics}
\label{sec:metrics}
\vspace{-1mm}
\noindent In order to perform quality assessment in video coding, two standard methodologies are usually employed: the computation of objective IQA/VQA metrics and subjective quality assessment. In this paper, we only perform objective quality assessment and leave the subjective evaluation, which is generally time-intensive, to be performed after we receive all participants' deliverables.\footnote{An extensive report of both objective and subjective quality assessment, as well as their correlations to the subjective scores will be presented during the special session in the conference.} Particularly, we explore the literature and evaluate commonly employed full reference and no reference IQA/VQA metrics. We decided to test both types of metrics due to the fact that: firstly, in video coding usually full-reference IQA/VQA metrics are used since the original video source is available (in most cases) and, secondly, in video denoising applications no reference metrics are typically used as there is no `reference' clean (denoised) sequence available.

\vspace{-2mm}
\subsection{Full Reference Metrics}
\vspace{-1mm}

\noindent Full Reference (FR) IQA/VQA metrics have been traditionally employed for video compression purposes and consist of a wide variety including Peak Signal to Noise Ratio (PSNR), that takes into account the Contrast Sensitivity Function (CSF) and the between-coefficient contrast masking of Discrete Cosine Transform (DCT) basis functions (PSNR-HVSM)~\cite{PSNR-HVSM}, Structural Similarity Index (SSIM)~\cite{Bovik_SSIM} and multi-scale SSIM (MS-SSIM)~\cite{Bovik_MSSSIM}, Visual Information Fidelity measure (VIF)~\cite{VIF} that is often employed as a feature,
Video Quality Metric (VQM)~\cite{VQM}, Spatio-Temporal Most Apparent Distortion model (ST-MAD)~\cite{STMAD} and Video Multi-method Assessment Fusion (VMAF) metric (using model vmaf\_v0.6.1.pkl, which was trained for FHD content)~\cite{VMAFblog}. PSNR, PSNR-HVSM, SSIM, MS-SSIM, and VIF are commonly used IQA metrics, while VQM, ST-MAD and VMAF are VQA methods. All of these metrics have their strengths and weaknesses in terms of their correlation to subjective scores and complexity. None of these metrics have been rigorously tested on low-light image or video compressed content. 
\vspace{-2mm}
\subsection{No Reference Metrics}
\vspace{-1mm}
\noindent No Reference (NR) IQA/VQA metrics are usually employed when the reference source sequences are not available (e.g.\ in user-generated content) or when the capture artefacts are dominant and the reference sequence is considered impaired. Several different no reference metrics have been proposed in the literature. 
The JPEG~\cite{Wang:NRJPEG:2002} and JPEG2000 (JP2K)~\cite{Sheikh:NRJPEG2000:2005} quality scores were two of the first no reference IQA methods introduced. The first attempts to align image
quality with HVS (Human Vision System) perception by characterising blockiness
and blurring. Since then a variety of different no reference quality metrics have been proposed in the literature, such as the Anisotropic Quality Index (AQI)~\cite{Gabarda:Blind:2007}, 
the Blind Image Quality Index (BIQI)~\cite{Moorthy:BIQI:2010}, the Contrast Enhancement (CEIQ) employed to measure contrast distortion  in~\cite{Fang:NRCD:2015}, the Naturaleness Image Quality Evaluator (NIQE)~\cite{NIQE}, the (PIQE)~\cite{PIQE}, the Blind/Referenceless Image Spatial Quality Evaluator (BRISQUE)~\cite{BRISQUE}, the Video BLIINDS~\cite{BLIINDS}, and the Two-Level Video Quality Metric (TL-VQM)~\cite{TLVQM}.  
Most of the aforementioned metrics have been designed based on Natural Scene Statistics theory, taking into account features such as contrast, intensity, colour, spatial and temporal correlation of frequencies and their statistical distributions. Lately, a lot of learning-based methods have been proposed, such as BRISQUE, TLVQM and more. However, none of these methods have been trained on low-light content.


\vspace{-1mm}
\section{Experiment Configurations}
\label{sec:methods}
\vspace{-1mm}
\subsection{Video Coding Conditions and Anchors}
\vspace{-1mm}
\noindent We followed one of the coding constraint cases in JVET Joint Call for Proposals on Video Compression with Capability beyond HEVC~\cite{s:beyondHEVC}:
\begin{itemize}[leftmargin=*]
	\item No more than 16 frames of structural delay, e.g.\ ``group of pictures'' (GOP) of 16.
	\item A random access interval of 1.1 seconds or less - defined as 32 pictures or less for a video sequence with a frame rate of 30 frames per second, 64 pictures or less for a video sequence with a frame rate of 64 frames per second.
\end{itemize}

\noindent We have selected a set of bitrates that are different for each sequence as shown in Table~\ref{tab:ResDO}. These target bitrates were selected after a small scale expert study, so that the perceived quality of the encoded anchor bitstreams will uniformly cover the quality scale from ``Poor'' up to ``Good'' and give to the participants room for improvement in the scale of perceived quality. 

 \begin{table}[t!]
 \centering
 	\caption{Test sequences and target bit rate points (in kbps).}
 	\label{tab:ResDO}
 	\footnotesize
 	\begin{tabular}{r l| rrrr}
 		\toprule
 		\multirow{2}{*}{No.}& \multirow{2}{*}{Sequences} &\multicolumn{4}{c}{\textbf{Target bit rates [kbps]}}  \\
 		\cmidrule{3-6}
 		 	&	& Rate1 & Rate2 & Rate3 & Rate4 	\\
 		\midrule S1  &ElFuente-Palacio& 100 & 170 & 300 &500  \\
 		\midrule S2  &ElFuente-Cars & 85	& 150&  280&  540  \\
 		\midrule S3  &ElFuente-Cyclist &70&	120 &  210&  400  \\
 		\midrule S4  &Chimera-Dinner & 50  &  70& 100 &200\\
 		\midrule S5  &Campfire	&  640 &1300  & 2500 &4500  \\
 		\midrule S6  &SmokeClear & 220	 &  400  & 700 &1400  \\
 		\bottomrule			
 	\end{tabular}
 \end{table}

\noindent Anchor bitstreams are generated using the VVC VTM 7.0 software, which uses static Quantisation Parameter (QP) configurations. A one-time change of QP, through the encoder parameter of \texttt{QPIncrementFrame (-qpif)} in VTM, was used to achieve the defined target bit rates for some bitstreams. The bitrates of produced anchor bitstreams do not exceed the target rate points by 3\%. We did not use rate control to hit the target bitrate as this compromises the quality of the resulting bitstream.

\vspace{-2mm}
\subsection{Benchmark methods}
\begin{figure}[t!]
	\centering
  		\includegraphics[width=\columnwidth]{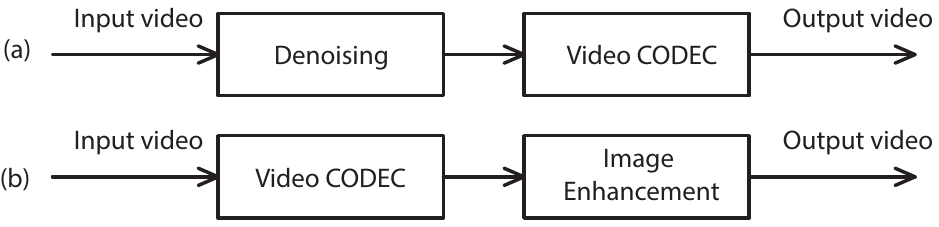}
  	\vspace{-.1cm}
	\caption{Diagrams of two processes for Encoding in the Dark: (a) preprocessing and (b) postprocessing methods.}
	\label{fig:diagram}
\end{figure}

\vspace{-1mm}
\noindent Two workflows examined and compared are shown in Fig~\ref{fig:diagram}. The preprocessing method applies denoising at the encoder, whilst the postprocessing method applied image enhancement at the decoder. Both of them operate on frame-by-frame basis.

 \vspace{-3mm}
\subsubsection{Preprocessing method with Denoising}
\label{sssec:preprocessing}
\vspace{-2mm}
\noindent Learning-based denoising methods have been proven to outperform traditional filtering techniques in both quality and speed. 
 Amongst these, Denoising CNN (DnCNN)~\cite{Zhang:DnCNN:2017} has become popular due to its reconstruction performance, simple implementation and  computational speed ~\cite{Lucas:Using:2018}.
The utilised DnCNN architecture comprises 17 convolutional layers combined with batch normalisation and a ReLU activation~\cite{Zhang:DnCNN:2017}. The network does not include pooling layers and therefore mainly extracts low-level features, fundamental for modelling noise in the image. In this paper, we used an adapted model trained on colour images with synthetic Gaussian noise.

We investigated the levels of temporal noise in denoised videos similar to that examined in Section~\ref{sec:dataset} shown in Fig.~\ref{fig:temporalsignals}. The MAEs ($\times$10$^{-3}$) of the nine points in the denoised sequence are  0.012,    1.678,    2.147,    2.167, 3.012,    3.869,    4.552,    4.697,    and 4.886, respectively. Eight of nine are less than the values of the original noisy videos, particularly in dark areas. This implies
that temporal variation is reduced after denoising, which results in more precise motion estimation with smaller residuals in the coding process.
\vspace{-3mm}
\subsubsection{Postprocessing Method with Image Enhancement}
\label{sssec:postprocessing}
\vspace{-2mm}
\noindent Postprocessing is commonly applied at the video decoder, on the reconstructed frames, to reduce various coding artefacts and enhance visual quality. Here, we employed the CNN-based postprocessing method proposed by Zhang et al.~\cite{Zhang:Enhancing:2020}, which has been reported to offer significant coding gains over VVC. Its network architecture was modified based on the generator (SRResNet) of SRGAN~\cite{Ledig:Photo:2017}. 
It contains 2$N$+2 convolutional layers, all of which have 3$\times$3 kernels, 64 feature maps and a stride value of 1, except the last convolutional layer (with 3 feature maps instead). Between the first and the last convolutional layers, there are $N$ identical residual blocks, each of which contains two convolutional layers and a parametric ReLU activation function in between them. Additional skip connections are employed (i)  between the input of the first residual block and the output of the $N^{\text{th}}$ residual block (ii) between the input of the CNN and the output of the last convolutional layer. Here we used $N$=16.
\vspace{-3mm}
\section{Results and discussion}
\label{sec:results}
\vspace{-1mm}
\noindent In the next subsections, we first discuss the results of the tested methods and then explain the limitations and challenges that will be part of our future work.
\vspace{-2mm}
\subsection{Results} 
\vspace{-1mm}
\noindent In the main phase of the evaluation, we executed the methods described in Section~\ref{sec:methods}.
The experiments were performed on a cluster computer, in which each node contains  2.4 GHz Broadwell Intel CPUs, 128GB RAM and 2x NVIDIA Tesla P100.

\begin{figure}[t!]
\scriptsize

\begin{minipage}[b]{0.47\linewidth}
\centering
\centerline{\includegraphics[width=1.02\linewidth]{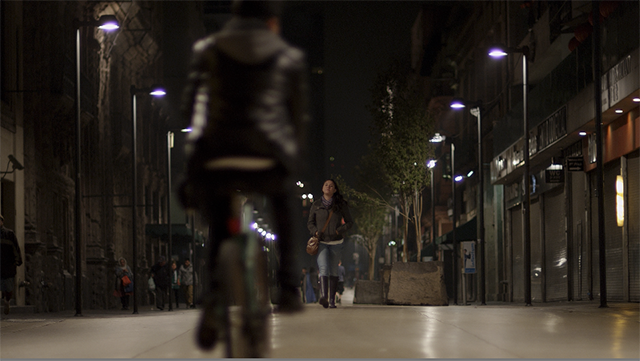}}
\centerline{(i) Denoised.}
\end{minipage}
\hfill
\begin{minipage}[b]{0.47\linewidth}
\centering
\centerline{\includegraphics[width=1.02\linewidth]{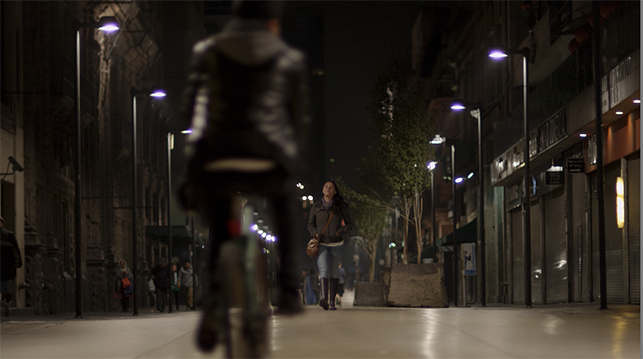}}
\centerline{(ii) VTM at QP=22.}
\end{minipage}

\begin{minipage}[b]{0.31\linewidth}
\centering
\centerline{\includegraphics[width=1.02\linewidth]{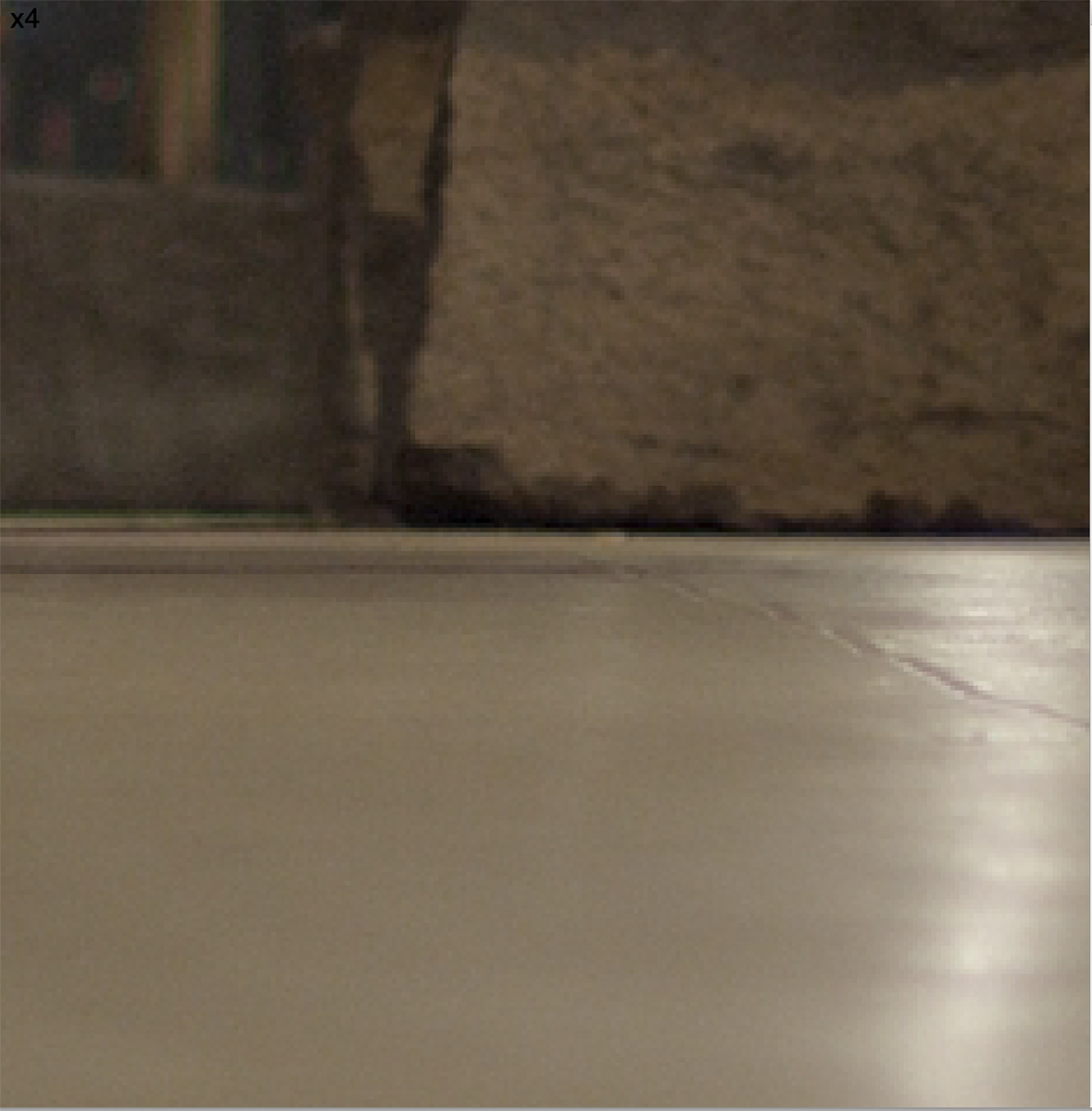}}
\centerline{(iii) Original.}
\end{minipage}
\hfill
\begin{minipage}[b]{0.31\linewidth}
\centering
\centerline{\includegraphics[width=1.02\linewidth]{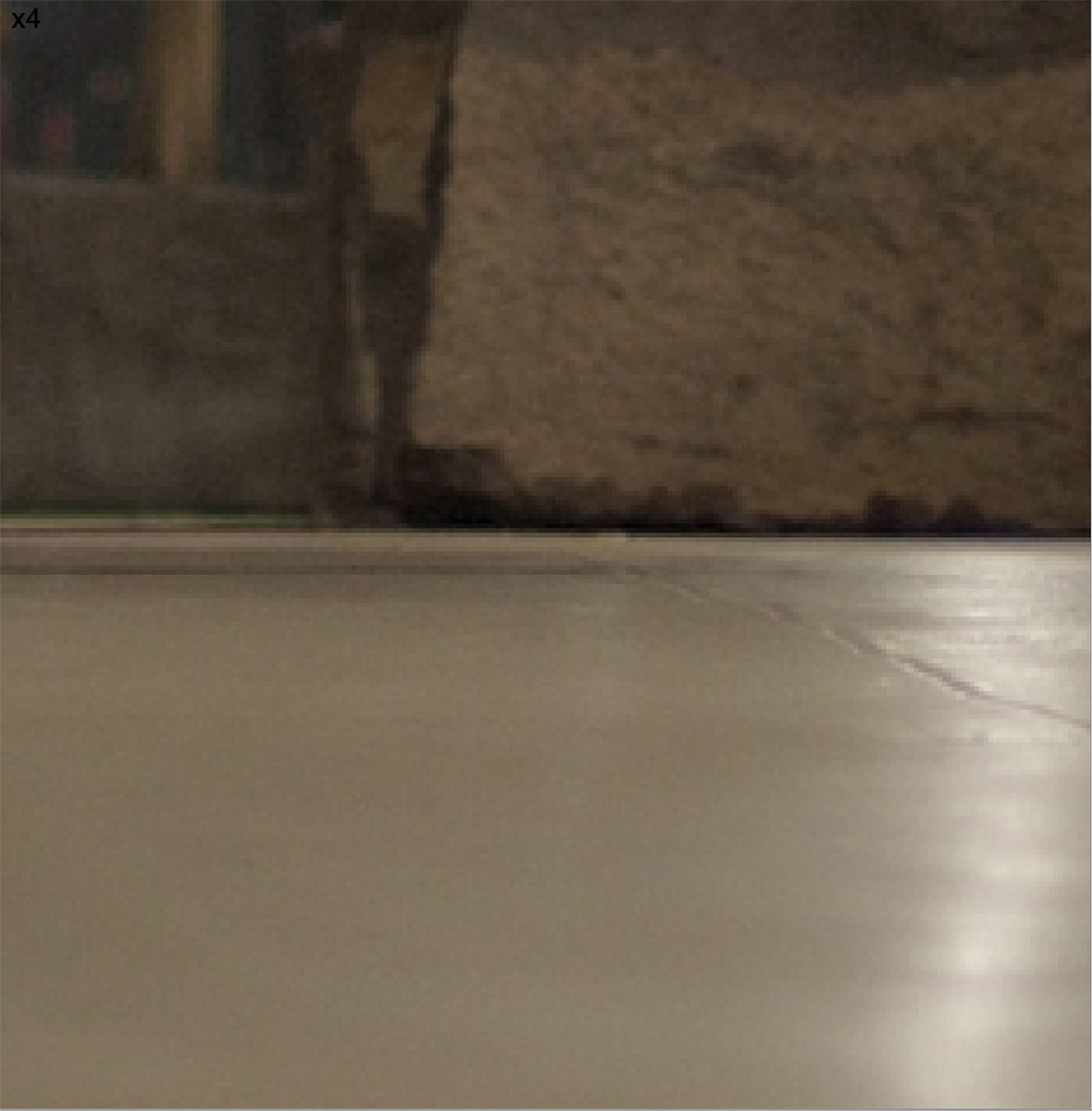}}
\centerline{(iv) Denoised.}
\end{minipage}
\hfill
\begin{minipage}[b]{0.31\linewidth}
\centering
\centerline{\includegraphics[width=1.02\linewidth]{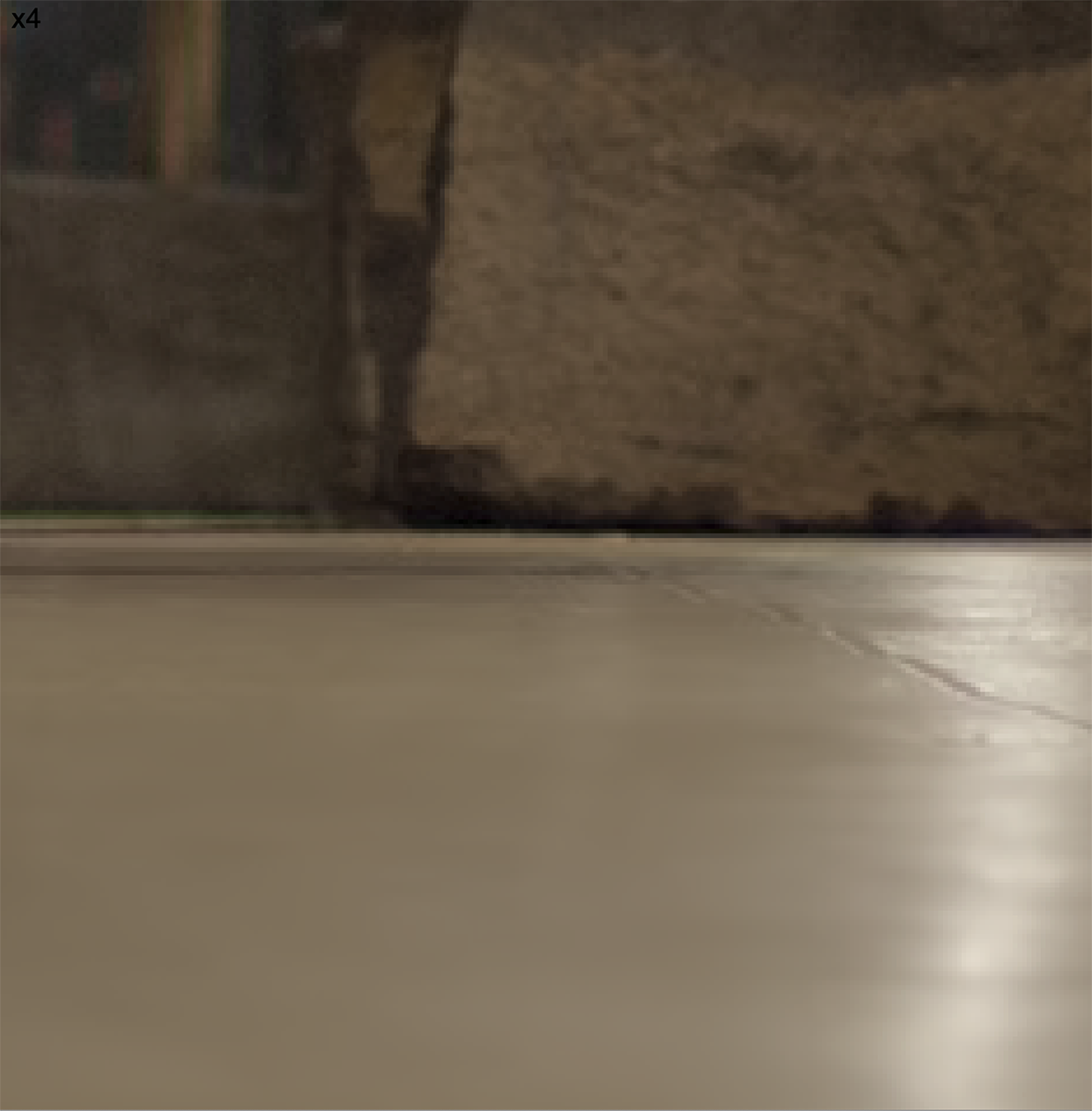}}
\centerline{(v) VTM at QP=22.}
\end{minipage}

\begin{minipage}[b]{0.47\linewidth}
\centering
\centerline{\includegraphics[width=1.02\linewidth]{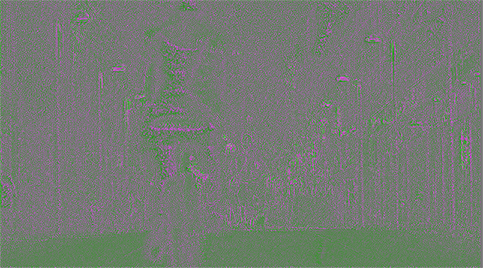}}
\centerline{(vi) Error of Original to Denoised.}
\end{minipage}
\hfill
\begin{minipage}[b]{0.47\linewidth}
\centering
\centerline{\includegraphics[width=1.02\linewidth]{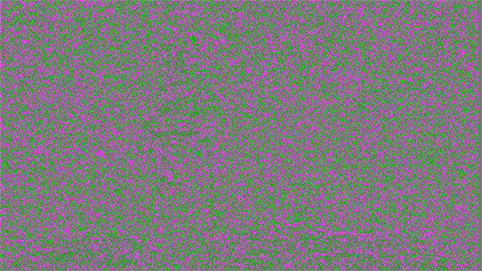}}
\centerline{(vii) Error of Original to VTM at QP=22.}
\end{minipage}
\vspace{-.1cm}

\caption{Example of visual inspection of S3 ElFuente-Cyclist sequence of the denoising effect by applying denoising or by just encoding with VVC.}
\vspace{-.3cm}
\label{fig:FrameDiff}
\end{figure}

First, we perform a visual inspection of the denoising effect that a deep-learning based method can provide against the denoising result of video encoding at light compression (at low QPs). In Fig.~\ref{fig:FrameDiff} (i)-(ii), we illustrate an example of the first frame of the S3 ElFuente-Cyclist sequence after applying denoising, as described in Section~\ref{sssec:preprocessing}, and the same frame after encoding the sequence with VTM7.0. Because the differences are almost imperceptible, we zoomed in at the same block and provide it at three different versions: in (iii) cropped from the original frame, in (iv) cropped from the denoised frame and in (v) cropped from the VTM encoded sequence. Figs.~\ref{fig:FrameDiff} (iii)-(iv) look very similar, while in (v) the compression effect is slightly apparent as the tiles look more uniform. The differences of the denoised and compressed frames from the respective original are confirmed by the visualisation of the frame difference in Figs.~\ref{fig:FrameDiff} (vi)-(vii). In these figures, the differences are amplified by a factor of 10. Green and pink colours represent pixel difference either positive or negative, and grey represents no difference. As can be seen, the differences are more intense for the compressed frame than the denoised.

From the resulting final YUV video sequences from the tested methods, we first plotted the rate-quality curves and then we computed the Bj{\o}ntegaard delta rate (BD-Rate)~\cite{r:Bjontegaard}. BD-Rate is widely used to calculate the coding efficiency between different coding technologies. The results of the pre- and post-processing methods are reported in Table~\ref{tab:BDRate}, where negative and positive values represent gain and loss of the coding performance, respectively. We computed the BD-Rate values for the different rate-quality curves, by taking into account the IQA/VQA metrics that indicated the best performance in terms of monotonicity. Particularly, we employed two FR metrics, PSNR of Y component (PSNR-Y) and VMAF, and two NR metrics, i.e. AQI and PIQE. PSNR is the most commonly used assessment method for video coding, while VMAF has been reported to offer better correlation with subjective opinions comparing to most of existing FR quality metrics \cite{zhang2018bvi}. Among most existing NF quality metrics, based on our preliminary study, the AQI and PIQE offer the best monotonicity characteristic against QP indices when coding the low-light sequences. Therefore these two metrics are employed here.

According to the PSNR-Y BD-Rate figures shown in Table~\ref{tab:BDRate}, the postprocessing method appears to improve the VTM performance with an average gain of 1.8\%. Conversely, applying denoising prior to encoding does not bring any gains but rather losses, with an average of 7.2\% in terms of PSNR. Inspecting the losses per sequence, the higher PSNR-Y BD-Rate losses are reported for sequences S4 and S5, which are the sequences with lower contrast and areas of very low luminosity. A similar conclusion can be drawn from the VMAF BD-Rate figures with slightly higher average gains/losses. 


Observing the NR BD-Rate figures in Table~\ref{tab:BDRate}, we notice that AQI shows a similar trend to the FR BD-Rate figures, but with amplified gains/losses. On the contrary, the PIQE BD-Rate figures are quite different, showing large differences in rate-quality curves of the tested methods against the anchors.

\begin{figure}[t!]
\scriptsize
\begin{minipage}[b]{0.45\linewidth}
\centering
\centerline{\includegraphics[width=\linewidth]{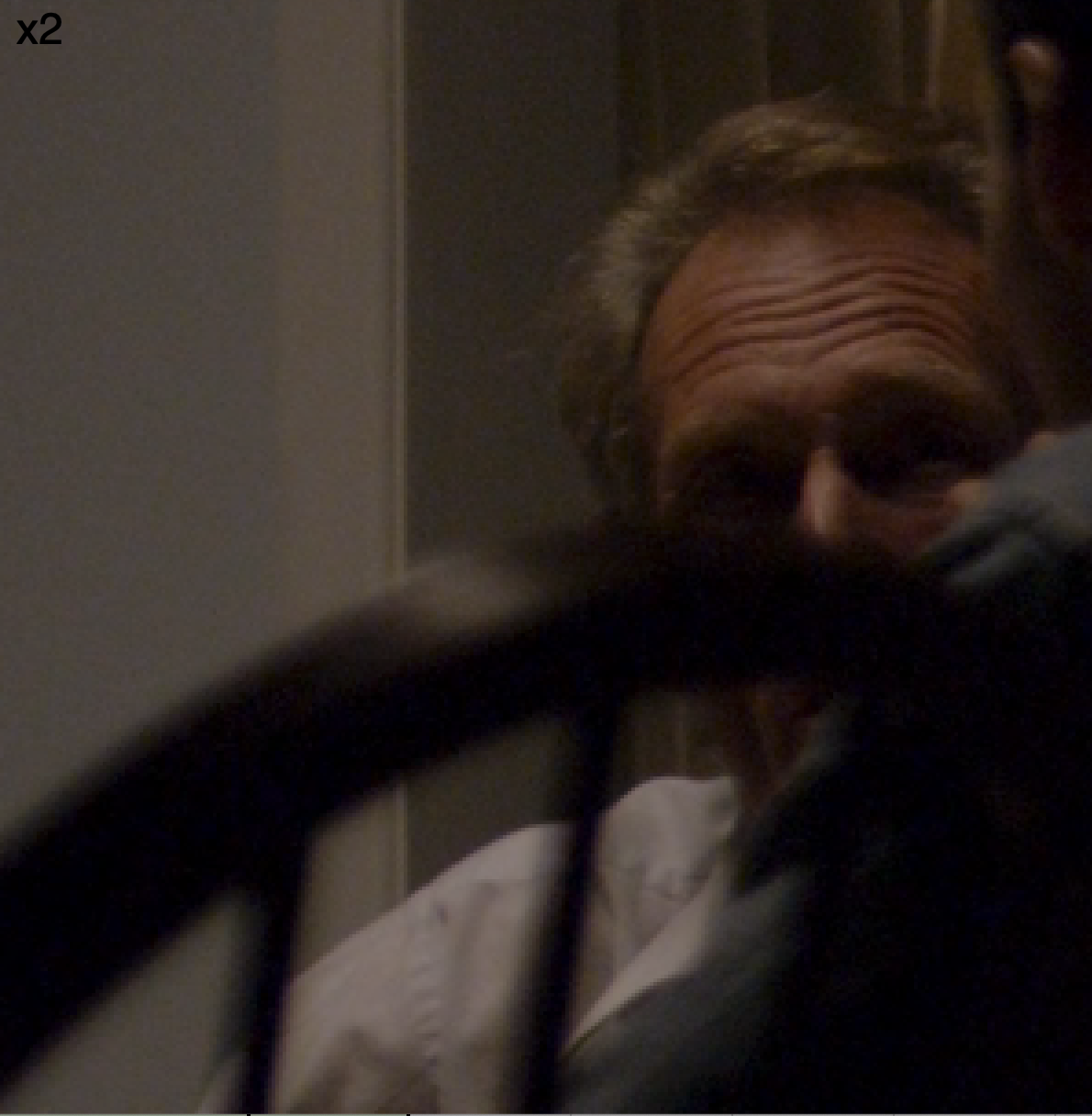}}
\centerline{(i) Original.}
\end{minipage}
\hfill
\begin{minipage}[b]{0.45\linewidth}
\centering
\centerline{\includegraphics[width=\linewidth]{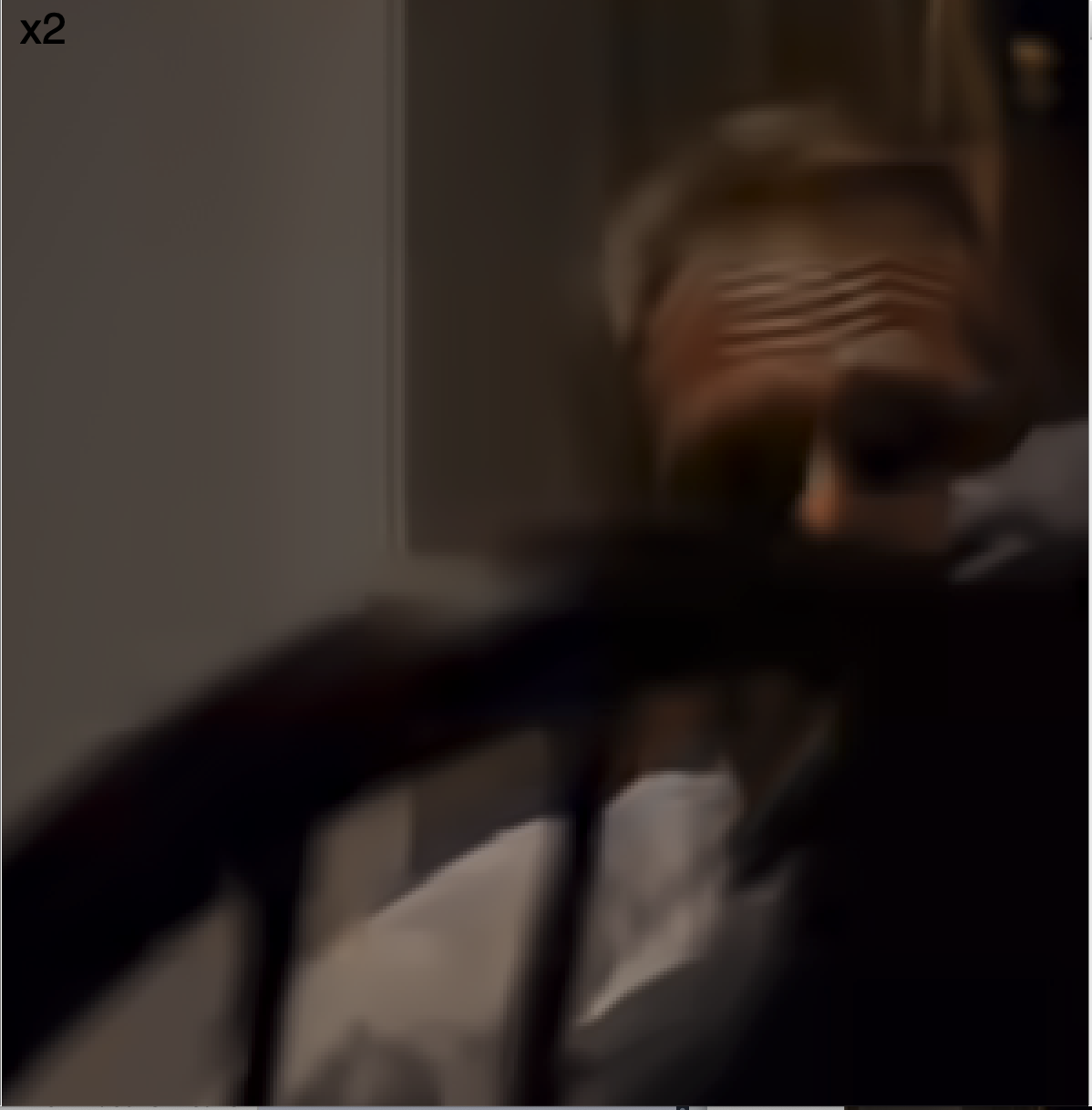}}
\centerline{(ii) Anchor.}
\end{minipage}
\hfill
\begin{minipage}[b]{0.45\linewidth}
\centering
\centerline{\includegraphics[width=\linewidth]{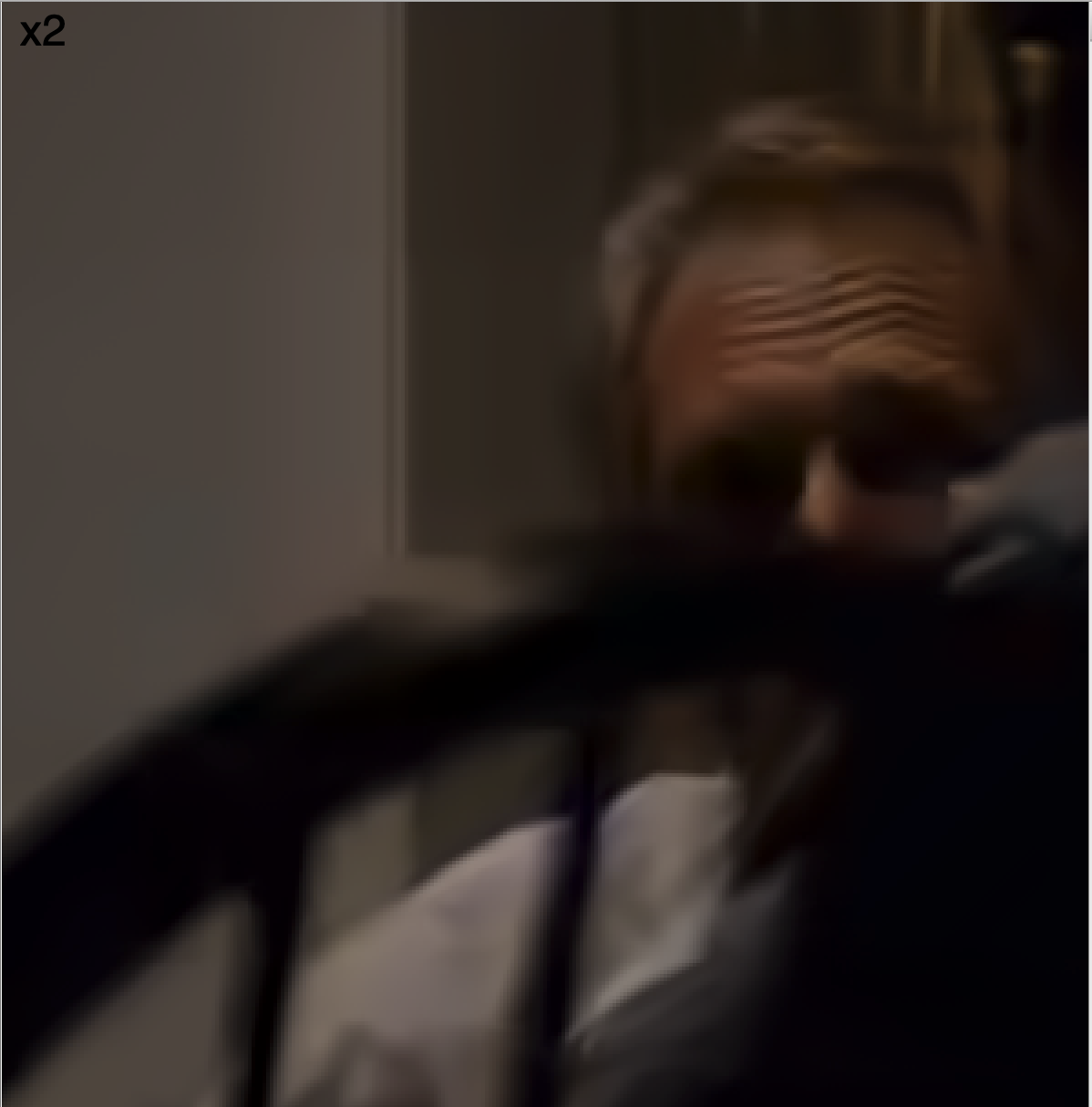}}
\centerline{(iii) Preprocessed.}
\end{minipage}
\hfill
\begin{minipage}[b]{0.45\linewidth}
\centering
\centerline{\includegraphics[width=\linewidth]{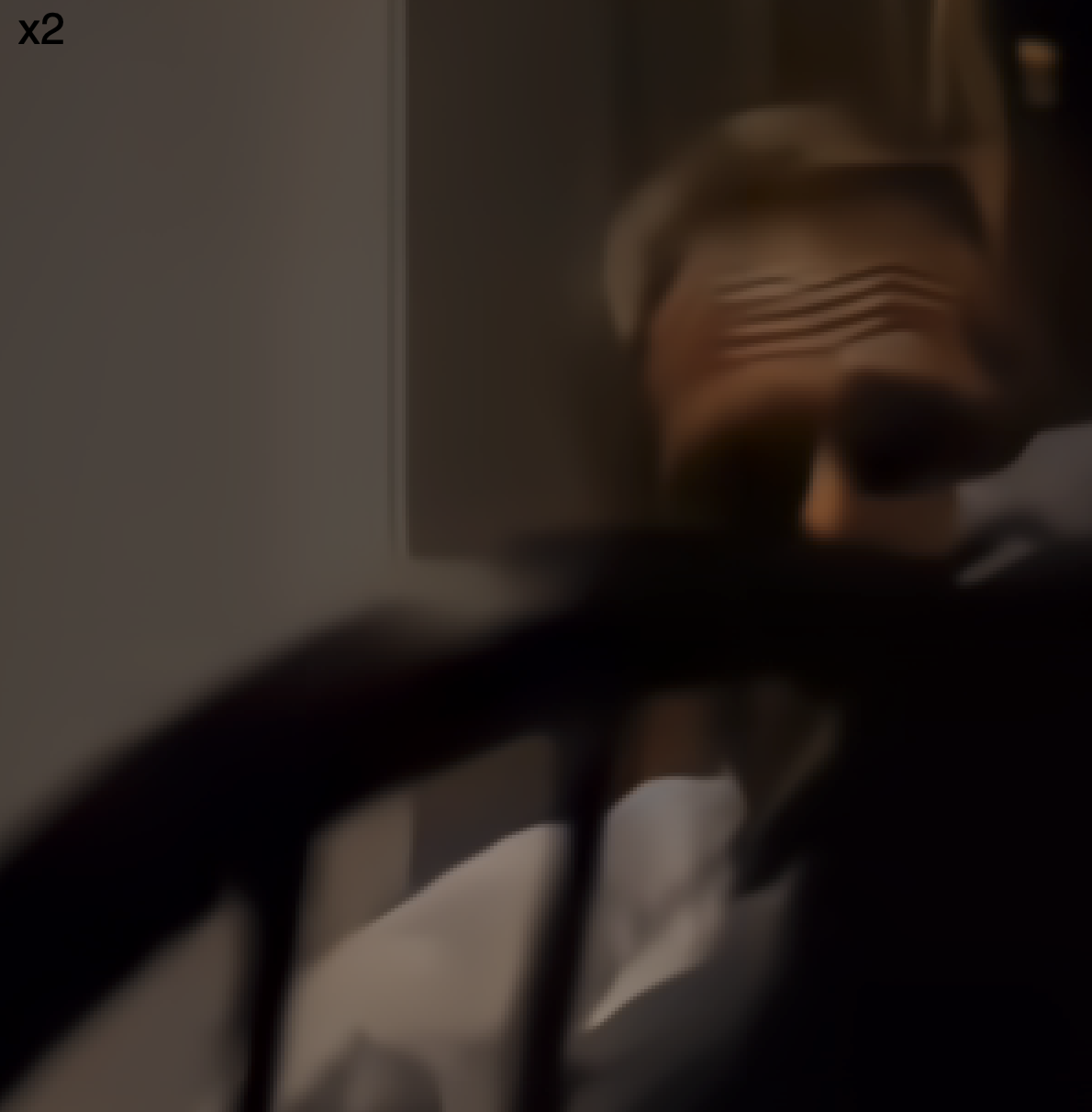}}
\centerline{(iv) Postprocessed.}
\end{minipage}
\vspace{-.2cm}

\caption{Example of visual inspection of S4 ElFuente-Dinner sequence comparing the tested methods at Rate 4.}
\label{fig:Crop}
\end{figure}

The above observations emphasize the need for a subjective quality assessment, so that we can confirm that the metrics utilised are suitable for dark scenes content. To further support this, we are illustrating in Fig.~\ref{fig:Crop} an example of cropped patches from sequence S4. As can be seen the quality is similar. The contrast in the preprocessed case is a bit higher and more details are preserved, for example the wrinkles in the forehead and the hair. In the case of the anchor and the postprocessed the forehead and the hair look flatter. All of these details though might not be noticeable during the video playout as they might be masked by motion. 

 \begin{table*}[t!]
\centering
 	\caption{BD-Rate savings achieved by two benchmark methods over VVC VTM 7.0, assessed by four different quality metrics~\cite{r:Bjontegaard}.}
 	\label{tab:BDRate}
 	\footnotesize
 	\begin{tabular}{c|c|cccccc|c}
 		\toprule
 		Metric & Method & S1 & S2 & S3 & S4 & S5 & S6 & Avg  \\
 		\midrule
 		 \multirow{2}{*}{PSNR$_{\text{Y}}$} & Pre & 1.1\% &	-0.5\%&	1.9\%&	8.0\%&	32.4\%& 0.33\%&7.2\%\\
 		 & Post & -3.0 \% &	-1.4\%&	-1.5\%&	-1.1\%	&-2.9\%	&-1.1\% &-1.8\%\\
 		 \midrule
 		 \multirow{2}{*}{VMAF}  & Pre & 3.5\%  &  0.0\%  &  6.3\%  &  8.6\% &  14.0\%  & -1.9\%&5.1\%\\
 		 & Post &  -5.7\%  & -5.9\% & -5.4\%  & -8.1\% &  -4.6\%  & 1.5\% &-4.7\%\\
 		\midrule
 		 \multirow{2}{*}{AQI} & Pre & 8.4\%&  -4.2\%&   -8.5\%  & 10.9\%&  -39.3\%&   -3.6\%& -6.0\%\\
 		 & Post & -9.5\% & -11.9\% & -14.9\%  &-20.5\%  & -10.0\%  &  2.6\%& -10.7\%\\
 		 \midrule
 		 \multirow{2}{*}{PIQE}	 & Pre & 77.0\% & 118.3\%&   61.5\% &  53.9\%  & 31.5\% & -52.3\%&48.3\%\\
 		 & Post &  -54.2\%  &-62.2\% & -58.2\%  &-21.5\% &  79.2\% & -46.8\%& -27.3\%\\
 		\bottomrule			
 	\end{tabular}
 \end{table*}

\vspace{-3mm}
\subsection{Limitations and Open Issues} 
\vspace{-1mm}
\noindent The results of the tested methods indicate some of the limitations and issues we encountered when we were preparing this Grand Challenge. The most important points can be summarised below:
\begin{itemize}[leftmargin=*]
    \item The lack of a large dataset with low-light video content certainly poses limitations on the effectiveness of learning-based methods. It is important to note that the models used in both pre- and post-processing frameworks are trained on generic datasets with natural scenes. 
    We expect that their performance would be improved, were they retrained using dark scenes.
    \item The absence of a dataset with subjective evaluations poses another limitation as the considered FR/NR IQA/VQA metrics can not be fully validated on their performance on this type of content. First of all, without subjective evaluation data, we cannot answer the critical question of which type of metric is more suitable for low-light sequences. On one hand, a denoised problem is considered as referenceless, while on the other hand in the video coding pipeline you always have a source/reference sequence.
    \item The previous point also leads to the question of whether the metrics examined here are the best performing ones and whether the BD-Rate gains/losses reported in this paper are representative. We anticipate that with the extensive subjective evaluation, to be conducted within this Challenge, we will re-evaluate all metrics and conclude on the most suitable for this type of content.
\end{itemize}

 \vspace{-2mm}
\section{Conclusion and Future work}
\label{sec:conclusion}
\vspace{-1mm}
\noindent This paper presents a study of encoding of low-light captured videos using contemporary methods, serving as the benchmark in the Grand Challenge within IEEE ICME2020. We carefully selected six dark-scene videos and provided anchors based on VVC. We then investigated available FR and NR evaluation metrics. Two workflows for encoding in the dark were examined: preprocessing with DnCNN-based denoising and postprocessing with a CNN-based image enhancement method. Experimental results show that the postprocessing framework outperforms the VTM alone by up to 1.8\% on average based on PSNR-Y, whilst the state-of-the-art VTM encoder appears to not improve its performance (7.2\%) by applying noise removal in the preprocessing framework.

In the future, the limitations and challenges identified will be further studied. Additional state-of-the-art methods will be benchmarked and compared against the Grand Challenge participants' deliverables. Furthermore, extensive experiments of subjective quality assessment will be performed to assess the perceptual gains of these methods. Another important outcome of this challenge is expected to be the evaluation of the correlation of the IQA/VQA methods with subjective quality for low-light sequences.
\vspace{-2mm}

\section{Acknowledgment}
The work was supported in part by the Bristol \& Bath R\&D cluster (AH/S002936/1) to N. Anantrasirichai, in part by the Leverhulme Early Career Fellowship (ECF-2017-413) awarded to A. Katsenou. We also gratefully acknowledge the support of Facebook and Netflix for sponsoring the awards of the finalists.
\small
\bibliographystyle{IEEEbib}
\vspace{-1mm}
\bibliography{icme2020_GC}

\begin{thebibliography}{10}

\bibitem{w:GameofThrones}
A.~Stout,
\newblock ``Game of thrones: Was the long night too dark?,''
\newblock {\em IBC}, May 2019.

\bibitem{Stankiewicz:video:2019}
O.~Stankiewicz,
\newblock ``Video coding technique with a parametric modelling of noise,''
\newblock {\em Opto-Electronics Review}, vol. 27, no. 3, pp. 241 -- 251, 2019.

\bibitem{KuoTCSVT2009}
B.~T. {Oh}, S.~{Lei}, and C.-J. {Kuo},
\newblock ``Advanced film grain noise extraction and synthesis for
  high-definition video coding,''
\newblock {\em IEEE Trans. Circ. Syst. Video Tech.}, vol. 19, no. 12, pp.
  1717--1729, Dec 2009.

\bibitem{Yahya:video:2016}
A.~A. {Yahya}, J.~{Tan}, B.~{Su}, and K.~{Liu},
\newblock ``Video denoising based on spatial-temporal filtering,''
\newblock in {\em 6th Intern. Conf. on Digital Home}, Dec 2016, pp. 34--37.

\bibitem{Malm:adaptive:2007}
H.~{Malm}, M.~{Oskarsson}, E.~{Warrant}, P.~{Clarberg}, J.~{Hasselgren}, and
  C.~{Lejdfors},
\newblock ``Adaptive enhancement and noise reduction in very low light-level
  video,''
\newblock in {\em IEEE ICCV}, Oct 2007, pp. 1--8.

\bibitem{Maggioni:BM4D:2012}
M.~Maggioni, V.~Katkovnik, K.~Egiazarian, and A.~Foi,
\newblock ``Nonlocal transform-domain filter for volumetric data denoising and
  reconstruction,''
\newblock {\em IEEE Trans. Image Process.}, vol. 22, no. 1, pp. 119--133, 2012.

\bibitem{Zuo:video:2013}
Zuo C, Liu Y, Tan X, Wang W, and Zhang M.,
\newblock ``Video denoising based on a spatiotemporal kalman-bilateral mixture
  model,''
\newblock {\em ScientificWorldJournal}, vol. 438147, 2013.

\bibitem{Buades:CFA:2019}
A.~{Buades} and J.~{Duran},
\newblock ``Cfa video denoising and demosaicking chain via spatio-temporal
  patch-based filtering,''
\newblock {\em IEEE Trans. Circ. Syst. Video Tech.}, pp. 1--1, 2019.

\bibitem{Zhang:DnCNN:2017}
K.~{Zhang}, W.~{Zuo}, Y.~{Chen}, D.~{Meng}, and L.~{Zhang},
\newblock ``Beyond a gaussian denoiser: Residual learning of deep cnn for image
  denoising,''
\newblock {\em IEEE Trans. on Image Processing}, vol. 26, no. 7, pp.
  3142--3155, July 2017.

\bibitem{Claus:ViDeNN:2019}
M.~{Claus} and J.~{van Gemert},
\newblock ``Videnn: Deep blind video denoising,''
\newblock in {\em CVPR workshop}, 2019.

\bibitem{Davy:nonlocal:2019}
A.~{Davy}, T.~{Ehret}, J.~{Morel}, P.~{Arias}, and G.~{Facciolo},
\newblock ``A non-local cnn for video denoising,''
\newblock in {\em IEEE ICIP}, Sep. 2019, pp. 2409--2413.

\bibitem{KaupCSVT2014}
E.~{Wige}, G.~{Yammine}, P.~{Amon}, A.~{Hutter}, and A.~{Kaup},
\newblock ``In-loop noise-filtered prediction for high efficiency video
  coding,''
\newblock {\em IEEE Trans. Circ. Syst. Video Tech.}, vol. 24, no. 7, pp.
  1142--1155, 2014.

\bibitem{YangICASSP2017}
M.~{Tang}, Y.~{Han}, J.~{Wen}, and S.~{Yang},
\newblock ``{HEVC-based motion compensated joint temporal-spatial video
  denoising},''
\newblock in {\em IEEE ICASSP}, March 2017, pp. 1797--1801.

\bibitem{JVET-O0079}
S.~{Wan}, M.-Z. {Wang}, H.~{Gong}, C.-Y. {Zou}, Y.-Z. {Ma}, J.-Y. {Huo}, Y.-F.
  {Yu}, and Y.~{Liu},
\newblock ``Ce10: Integrated in- loop filter based on cnn (tests 2.1, 2.2 and
  2.3),''
\newblock Tech. {R}ep. {JVET meeting, no. JVET-O0079. ITU-T, ISO/IEC}, ITU-R,
  2019.

\bibitem{MaAccess2019}
M.~{Wang}, S.~{Wan}, H.~{Gong}, and M.~{Ma},
\newblock ``{Attention-Based Dual-Scale CNN In-Loop Filter for Versatile Video
  Coding},''
\newblock {\em IEEE Access}, vol. 7, pp. 145214--145226, 2019.

\bibitem{Zhang:Enhancing:2020}
F.~{Zhang}, F.~{Chen}, and D.~R. {Bull},
\newblock ``{Enhancing VVC through CNN-based Post-Processing},''
\newblock in {\em IEEE ICME}, 2020.

\bibitem{w:NetflixElFuente}
I~Katsavounidis,
\newblock ``{NETFLIX - ``El Fuente''} video sequence details and scenes,'' July
  2015.

\bibitem{w:NetflixChimera}
I~Katsavounidis,
\newblock ``{NETFLIX - ``Chimera''} video sequence details and scenes,''
  November 2015.

\bibitem{c:SJTU4K}
Y.~Zhu, L.~Song, R.~Xie, and W.~Zhang,
\newblock ``{SJTU} {4K} video subjective quality dataset for content adaptive
  bit rate estimation without encoding,''
\newblock in {\em Broadband Multimedia Systems and Broadcasting (BMSB), 2016
  IEEE Intern. Symposium on}. IEEE, 2016, pp. 1--4.

\bibitem{Papadopoulos2015}
M.~A. Papadopoulos, F.~Zhang, D.~Agrafiotis, and D.~Bull,
\newblock ``A video texture database for perceptual compression and quality
  assessment,''
\newblock in {\em IEEE ICIP}, 2015, pp. 2781--2785.

\bibitem{r:h264}
{ITU-T Rec. H.264},
\newblock ``Advanced video coding for generic audiovisual services,'' 2005.

\bibitem{r:HEVC}
{ITU-T Rec. H.265},
\newblock ``High efficiency video coding,'' 2015.

\bibitem{b:Wien}
M.~Wien,
\newblock {\em High efficiency video coding},
\newblock Springer, 2015.

\bibitem{j:Ohm}
J.~R. Ohm, G.~J. Sullivan, H.~Schwarz, T.~K. Tan, and T.~Wiegand,
\newblock ``Comparison of the coding efficiency of video coding standard -
  including {H}igh {E}fficiency {V}ideo {C}oding ({HEVC}),''
\newblock {\em IEEE Trans. Circ. Syst. Video Tech.}, vol. 22, no. 12, pp.
  1669--1684, 2012.

\bibitem{w:AV1}
{AOM},
\newblock ``{AOMedia Video 1 (AV1)},'' \url{https://github.com/AOMediaCodec},
  2019.

\bibitem{w:VP9}
``{VP9 Video Codec},'' \url{https://www.webmproject.org/vp9/}.

\bibitem{c:Zhang24}
A.~V. Katsenou, F.~Zhang, M.~Afonso, and D.~R. Bull,
\newblock ``A subjective comparison of {AV1} and {HEVC} for adaptive video
  streaming,''
\newblock in {\em IEEE ICIP}, 2019.

\bibitem{c:Topiwala}
P.~Topiwala, M.~Krishnan, and W.~Dai,
\newblock ``Performance comparison of vvc, av1 and hevc on 8-bit and 10-bit
  content,''
\newblock in {\em Applications of Digital Image Processing XLI}. Intern.
  Society for Optics and Photonics, 2018, vol. 10752, p. 107520V.

\bibitem{c:Grois}
D.~Grois, T.~Nguyen, and D.~Marpe,
\newblock ``Coding efficiency comparison of {AV1, VP9, H.265/MPEG-HEVC, and H.
  264/MPEG-AVC} encoders,''
\newblock in {\em Picture Coding Symposium (PCS)}. IEEE, 2016, pp. 1--5.

\bibitem{c:Nguyen1}
T.~Nguyen and D.~Marpe,
\newblock ``Future video coding technologies: A performance evaluation of av1,
  jem, vp9, and hm,''
\newblock in {\em 2018 Picture Coding Symposium (PCS)}. IEEE, 2018, pp. 31--35.

\bibitem{PSNR-HVSM}
N.~Ponomarenko, F.~Silvestri, K.~Egiazarian, M.~Carli, J.~Astola, and V.~Lukin,
\newblock ``{On between-coefficient contrast masking of DCT basis functions},''
\newblock in {\em Proc. of the 3rd Intern. workshop on video processing and
  quality metrics}, 2007, vol.~4.

\bibitem{Bovik_SSIM}
Z.~{Wang}, A.~C. {Bovik}, H.~R. {Sheikh}, and E.~P. {Simoncelli},
\newblock ``Image quality assessment: from error visibility to structural
  similarity,''
\newblock {\em IEEE Trans. Image Process.}, vol. 13, no. 4, pp. 600--612, April
  2004.

\bibitem{Bovik_MSSSIM}
Z.~Wang, E.~P. Simoncelli, and A.~C. Bovik,
\newblock ``Multi-scale structural similarity for image quality assessment,''
\newblock in {\em Asilomar Conf. Signals Syst. Comput.}, 2003, pp. 1398--1402.

\bibitem{VIF}
H.~R. {Sheikh}, A.~C. {Bovik}, and G.~{de Veciana},
\newblock ``An information fidelity criterion for image quality assessment
  using natural scene statistics,''
\newblock {\em IEEE Trans. Image Process.}, vol. 14, no. 12, pp. 2117--2128,
  Dec 2005.

\bibitem{VQM}
M.~H. {Pinson} and S.~{Wolf},
\newblock ``A new standardized method for objectively measuring video
  quality,''
\newblock {\em IEEE Trans. on Broadcasting}, vol. 50, no. 3, pp. 312--322, Sep.
  2004.

\bibitem{STMAD}
P.~V. {Vu}, C.~T. {Vu}, and D.~M. {Chandler},
\newblock ``A spatiotemporal most-apparent-distortion model for video quality
  assessment,''
\newblock in {\em IEEE ICIP}, Sep. 2011, pp. 2505--2508.

\bibitem{VMAFblog}
Z.~Li, A.~Aaron, I.~Katsavounidis, A.~Moorthy, and M.~Manohara,
\newblock ``{The NETFLIX tech blog: Toward a practical perceptual video quality
  metric},''
  \url{http://techblog.netflix.com/2016/06/toward-practical-perceptual-video.html},
  note = {[Online; accessed 2018-08-04]},.

\bibitem{Wang:NRJPEG:2002}
Z.~Wang, H.~Sheikh, and A.~Bovik,
\newblock ``No-reference perceptual quality assessment of jpeg compressed
  images,''
\newblock in {\em IEEE ICIP}, 2002, pp. 477--480.

\bibitem{Sheikh:NRJPEG2000:2005}
H.~Sheikh, A.~Bovik, and L.~Cormack,
\newblock ``No-reference perceptual quality assessment of jpeg compressed
  images,''
\newblock {\em IEEE Trans. Image Process.}, vol. 14, no. 11, pp. 1918--1927,
  2002.

\bibitem{Gabarda:Blind:2007}
S.~Gabarda and G.~Cristobal,
\newblock ``Blind image quality assessment through anisotropy,''
\newblock {\em J. Opt. Soc. Am}, vol. 24, no. 12, pp. B42--B51, 2007.

\bibitem{Moorthy:BIQI:2010}
A.~K. {Moorthy} and A.~C. {Bovik},
\newblock ``A two-step framework for constructing blind image quality
  indices,''
\newblock {\em IEEE Signal Process. Lett.}, vol. 17, no. 5, pp. 513--516, 2010.

\bibitem{Fang:NRCD:2015}
Y.~{Fang}, K.~{Ma}, Z.~{Wang}, W.~{Lin}, Z.~{Fang}, and G.~{Zhai},
\newblock ``No-reference quality assessment of contrast-distorted images based
  on natural scene statistics,''
\newblock {\em IEEE Signal Process. Lett.}, vol. 22, no. 7, pp. 838--842, 2015.

\bibitem{NIQE}
A.~{Mittal}, R.~{Soundararajan}, and A.~C. {Bovik},
\newblock ``Making a “completely blind” image quality analyzer,''
\newblock {\em IEEE Signal Process. Lett.}, vol. 20, no. 3, pp. 209--212, 2013.

\bibitem{PIQE}
A.~K. {Moorthy} and A.~C. {Bovik},
\newblock ``Blind image quality assessment: From natural scene statistics to
  perceptual quality,''
\newblock {\em IEEE Trans. Image Process.}, vol. 20, no. 12, pp. 3350--3364,
  2011.

\bibitem{BRISQUE}
A.~{Mittal}, A.~K. {Moorthy}, and A.~C. {Bovik},
\newblock ``No-reference image quality assessment in the spatial domain,''
\newblock {\em IEEE Trans. Image Process.}, vol. 21, no. 12, pp. 4695--4708,
  2012.

\bibitem{BLIINDS}
M.~A. {Saad}, A.~C. {Bovik}, and C.~{Charrier},
\newblock ``Blind prediction of natural video quality,''
\newblock {\em IEEE Trans. Image Process.}, vol. 23, no. 3, pp. 1352--1365,
  2014.

\bibitem{TLVQM}
J.~{Korhonen},
\newblock ``{Two-Level Approach for No-Reference Consumer Video Quality
  Assessment},''
\newblock {\em IEEE Trans. Image Process.}, vol. 28, no. 12, pp. 5923--5938,
  2019.

\bibitem{s:beyondHEVC}
A.~Segall, V.~Baroncini, J.~Boyce, J.~Chen, and T.~Suzuki,
\newblock ``Joint call for proposals on video compression with capability
  beyond hevc,'' October 2017.

\bibitem{Lucas:Using:2018}
A.~Lucas, M.~Iliadis, R.~Molina, and A.~K. Katsaggelos,
\newblock ``Using deep neural networks for inverse problems in imaging: beyond
  analytical methods,''
\newblock {\em IEEE Signal Processing Magazine}, vol. 35, no. 1, pp. 20--36,
  2018.

\bibitem{Ledig:Photo:2017}
C.~Ledig and et~al.,
\newblock ``Photo-realistic single image super- resolution using a generative
  adversarial network,''
\newblock in {\em IEEE CVPR}, 2017, p. 105–114.

\bibitem{r:Bjontegaard}
G.~Bj{\o}ntegaard,
\newblock ``Calculation of average {PSNR} differences between {RD}-curves,''
  April 2001.

\bibitem{zhang2018bvi}
Fan Zhang, Felix~Mercer Moss, Roland Baddeley, and David~R Bull,
\newblock ``{BVI-HD}: A video quality database for {HEVC} compressed and
  texture synthesized content,''
\newblock {\em IEEE Transactions on Multimedia}, vol. 20, no. 10, pp.
  2620--2630, 2018.

\end{thebibliography}
\end{document}